\documentclass[12pt]{article}

\usepackage{amsthm,graphicx,cite,lscape,float}
\usepackage{epsf,latexsym}
\usepackage{amssymb,amsmath}

\usepackage{epsf,latexsym}
\usepackage{amssymb,amsmath,caption}
\newcommand{\nhat}{\hat{\rm n}}
\newcommand{\xhat}{\hat{\rm x}}
\newcommand{\fp}{\vec{f}_{\mbox{\scriptsize $\parallel$}}}
\newcommand{\fpm}{f_{\mbox{\scriptsize $\parallel$}}}

\begin{document}

\title{\bf Comparison of the lateral retention forces on sessile,
pendant, and inverted sessile drops}

\author{Rafael de la 
Madrid,\footnote{E-mail: \texttt{rafael.delamadrid@lamar.edu}} \ 
Fabian Garza,\footnote{Current address: Department of Physics, 
Texas A\&M University, College Station, TX 77843} \
Justin Kirk, 
Huy Luong,\footnote{Current address: Sage Automation, Beaumont, TX 77705} \\
Levi Snowden, 
Jonathan Taylor,$^{\dagger}$ 
Benjamin Vizena\footnote{Current address: METECS, Houston, TX 77289} \\ [2ex]
\small{\it Department of Physics, Lamar University,
Beaumont, TX 77710} }

\date{\small{\today}}


\maketitle

\begin{abstract}

\noindent We compare the lateral retention forces on sessile drops (which
are drops that are placed on top of a solid surface), pendant drops
(which are drops that are placed on the underside of the surface), and
inverted sessile drops (which are drops that are first placed on top and then 
on the underside of the surface by flipping the surface). We have found 
experimentally that the 
retention force on a truly pendant drop is always smaller than that on 
a sessile drop. However, the retention force on an inverted sessile drop 
is comparable to, and usually larger than, that on a sessile 
drop. Thus, the retention force on a drop depends not only on whether it
is placed on top or on bottom of a surface, but also on the history
of drop deposition, since such history affects the width, the shape and the
contact angles of the drop.
\end{abstract}

\vskip0.5cm



\newpage

\section{Introduction}

The study of liquid drops on solid substrates has attracted a great deal
of attention~\cite{DEGENNES,ERBIL,BORMASHENKO}, because of both its 
scientific interest and its industrial applications. However, there are 
many aspects of wetting and dewetting phenomena that are 
still not well understood. One such aspect was reported in 
Ref.~\cite{TADMOR09}, where Tadmor {\it et al.}~presented a very 
counterintuitive property of the lateral retention force on liquid droplets
at the moment the droplets start to slide on a solid surface. Contrary to 
the solid-solid friction case, a drop hanging from a solid surface 
experiences a larger retention force than a drop resting on the 
surface~\cite{TADMOR09}. In spite of the attention it drew~\cite{FOCUS,PT}, 
the origin of this effect remains unknown.

The purpose of this paper is to put forward an explanation of the effect
observed in Ref.~\cite{TADMOR09}. We will argue that the origin of 
such effect is rooted on how the drops are formed. We will distinguish 
between {\it sessile} drops (which are drops that are formed by placing a small
amount of liquid on top of a uniform, flat, solid surface),
{\it pendant} drops (which are drops that are formed by placing the 
liquid on the underside of the surface), and
{\it inverted sessile} (or simply {\it inverted}) drops (which are drops 
that are formed by placing the liquid on top of the 
surface, and then placing it on the underside 
by flipping the surface). We have found that the retention force on 
a truly pendant drop is smaller than that on a sessile drop, just as naive 
intuition suggests. We have also found that the retention 
force on an inverted sessile drop is comparable to, and usually larger 
than, that on a sessile drop. This
result may explain the effect observed in Ref.~\cite{TADMOR09}, 
because the drops placed on the underside
of the solid substrate in Ref.~\cite{TADMOR09} were not truly pendant drops,
but rather inverted sessile drops.

\section{Experimental Section}

\begin{figure}[h!]
\begin{center}
              \epsfxsize=8cm
              \epsffile{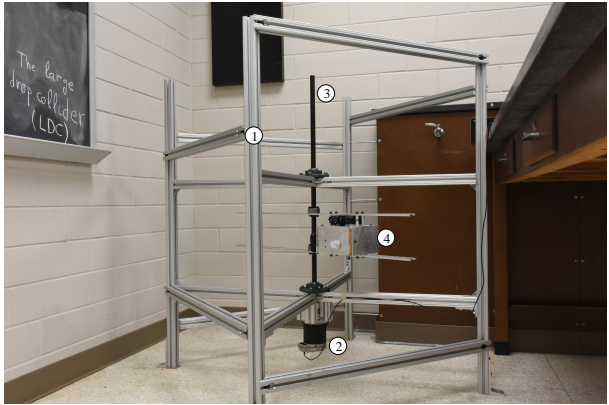}
\end{center}                
\caption{Drop accelerator: {\bf 1}-aluminum frame; {\bf 2}-motor; 
{\bf 3}-shaft; {\bf 4}-box.}
\label{fig:dropa}
          \end{figure}

Our experimental apparatus is a simplified version of the Centrifugal
Adhesion Balance~\cite{TADMOR09} and the Kerberos drop 
accelerator~\cite{GRIEGOS1,GRIEGOS2,GRIEGOS3}. It consists of a metallic 
frame with dimensions
$90~\text{cm} \times 90~\text{cm} \times 120~\text{cm}$ inside of which 
the rotary unit is mounted, 
see Fig.~\ref{fig:dropa}. The four legs of the frame are 
bolted to the floor to reduce mechanical vibrations. The rotary unit 
consists of a servo motor~\cite{TEKNIC}, a shaft, and two
pairs of aluminum rails that are attached perpendicularly to the shaft. The 
motor is connected to a power supply and a computer, whose software 
controls the motor.

\begin{figure}[h!]
\begin{center}
             \epsfxsize=8cm
             \epsffile{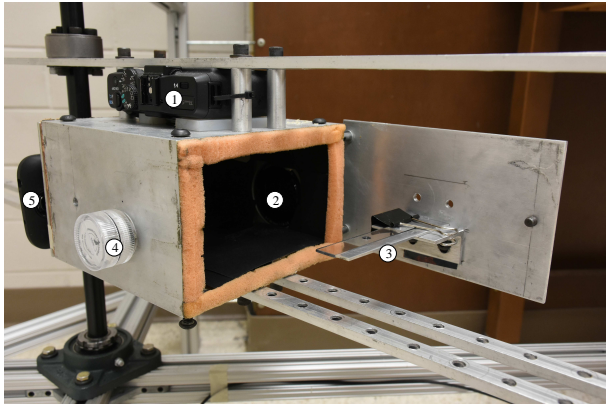}
\end{center}                
\caption{Box: {\bf 1}-top camera; {\bf 2}-side camera;
{\bf 3}-PMMA sheet with drop; {\bf 4}-remote-controlled LED light;
{\bf 5}-smart phone with accelerometer. Inside 
the box (not shown
in the picture) there is an LED panel that provides the necessary illumination
for the top camera.}
\label{fig:box}
          \end{figure}

A metallic box is mounted on the aluminum rails. The box
has two cameras placed on top and on the side, which
provide top and side views of the drops, see Fig.~\ref{fig:box}. A 
remote-controlled LED is used to set a common starting time in
the videos of the cameras. On the door of the metallic box, we mounted a 
poly methyl methacrylate (PMMA) sheet (Optix$^{\tiny \textregistered}$, by 
Plaskolite) such that, when a drop is placed on 
the sheet and the door is closed, the cameras have side and top views of the 
drop. Illumination for
the side camera is provided by the remote-controlled LED. Lighting for the
top camera is provided by an LED panel and an optical 
gradient~\cite{PODGORSKI}. To remove any remnants of their protective 
films, the PMMA sheets were initially washed with hot water and soap. 
Afterward, before each run, the PMMA sheets were cleaned
with 70\% isopropyl alcohol and paper tissue~\cite{AJP15}, and dried 
with lamplight.

\begin{figure}[h!]
\begin{center}
              \epsfxsize=5.5cm
              \epsffile{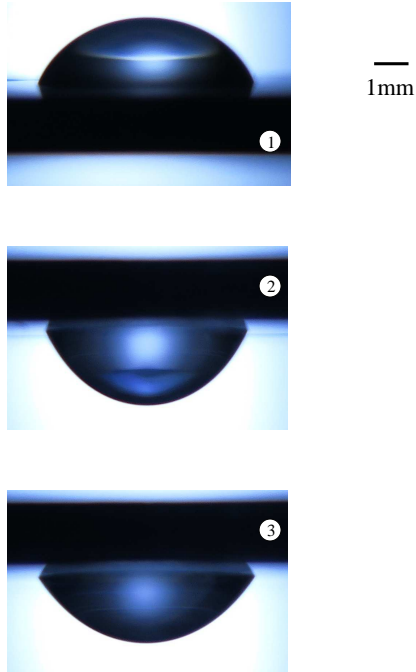}
\end{center}                
\caption{Side view of 30-$\mu$L sessile {\bf 1}, pendant {\bf 2}, and 
inverted sessile {\bf 3} drops at rest.}
\label{fig:comparison}
\end{figure}

To place the water droplets on the PMMA sheet, we used
a micropipette~\cite{SYRINGE}. A sessile (pendant) drop was 
formed by slowly releasing the contents of the micropipette on top (bottom) 
of the PMMA sheet. An inverted sessile drop was formed by slowly 
releasing the contents of the micropipette on top of the PMMA 
sheet, and then flipping the sheet, so the drop ends up on the underside of the 
sheet. Figure~\ref{fig:comparison} shows 30-$\mu$L sessile, pendant, and
inverted sessile drops at rest. Clearly, an inverted sessile drop is not 
the same as a pendant drop. In particular, the width of the inverted sessile
drop is larger than that of a pendant drop, but comparable
to the width of the sessile drop. As we will see, the width
is the main factor that differentiates the retention forces on 
sessile, pendant, and inverted sessile drops.

To make the drops slide on the PMMA sheet, we rotated the box 
with a constant angular acceleration of about 0.086~rad/s$^2$. As the 
angular velocity of the drop increases, the centrifugal force
increases~\cite{VIBRATION}. This eventually makes the advancing edge of 
the drop crawl 
forward in the radial direction, although the receding edge of the drop stays 
pinned to the surface~\cite{GRIEGOS2}. While the advancing edge crawls 
forward, the triple line deforms slightly from its initial circular shape. At 
some point, when the centrifugal force is large enough, the receding edge of 
the drop also starts moving in the outward, radial direction. When
that happens, the whole drop moves in the outward, radial 
direction~\cite{GRIEGOS2}. This is why
we identify the onset of the motion of the drop with the instant at which the 
receding (i.e., trailing) edge of the drop starts moving. 

We used the videos of the side camera and custom-made software to 
determine the instant when the receding edge of the drop starts 
moving. At such instant, we obtained the contact angles from 
the videos of the side-view camera using ImageJ~\cite{IMAGEJ}, and the width
from the videos of the top-view camera using PixelZoomer~\cite{PZ}.

\section{Results and Discussion}

\subsection{Experimental Results}

We measured the time it took each drop to start sliding on the
PMMA sheet for each type of drop
(sessile, pendant, and inverted sessile), for each volume
(15, 20, 25, and 30~$\mu$L), and for twelve different PMMA sheets. For 
volumes larger than 30~$\mu$L, it is hard to produce inverted
drops, because big drops slide on the surface during the flip. Hence, 
for 40-100~$\mu$L volumes, we analyzed only sessile and pendant drops at 
intervals of 10~$\mu$L. Using the times at which the onset of the
motion occurs, we obtained the lateral retention force from the centrifugal 
force,
\begin{equation}
         F = m r \omega ^2 = \rho V r \alpha ^2 t^2 \, ,
\end{equation}
where $m$ is the mass of the drop, $\rho$ is its density, $V$ is its 
volume, $r$ is the distance from the center of the drop to the axis of
rotation (about 170~mm in our experiment), $\alpha$ is the angular 
acceleration, and $t$ is the time since the rotation of the motor started.

For each volume and for each type of drop, we calculated the average time
of about 24 runs, and we obtained that
\begin{equation}
         t_{\rm pendant}<t_{\rm sessile} <t_{\rm inverted} \, .
\end{equation}
Hence,
\begin{equation}
         F_{\rm pendant}<F_{\rm sessile} <F_{\rm inverted} \, .
          \label{seqoffor}
\end{equation}

\begin{figure}[h!]
\begin{center}
              \epsfxsize=8cm
              \epsffile{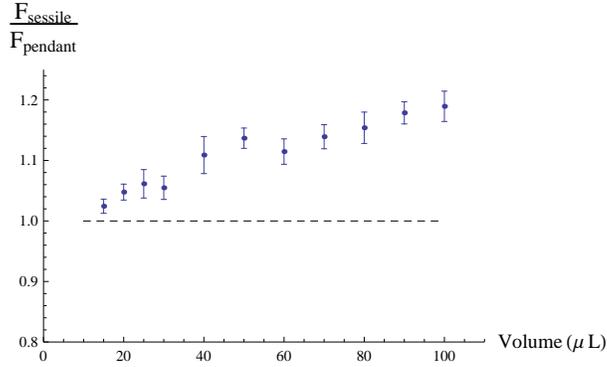}
\end{center}                
\caption{Ratio $\frac{F_{\rm sessile}}{F_{\rm pendant}}$
for 15-100~$\mu$L drops. Error bars are only statistical. The 
dashed line represents unity.}
\label{fig:sop}
          \end{figure}

\begin{figure}[h!]
\begin{center}
              \epsfxsize=8cm
              \epsffile{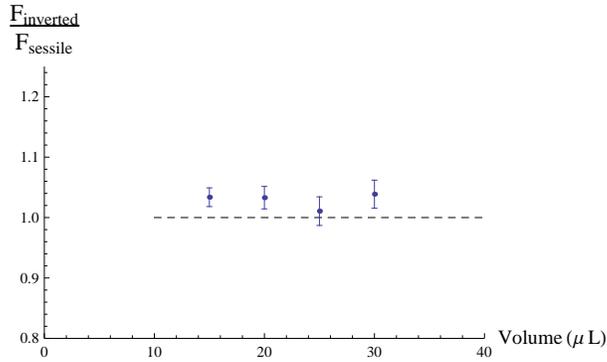}
\end{center}                
\caption{Ratio $\frac{F_{\rm inverted}}{F_{\rm sessile}}$ for 15-30~$\mu$L 
drops. Error bars are only statistical. The dashed line represents
unity.}
\label{fig:ios}
\end{figure}

The best way to visualize Eq.~(\ref{seqoffor}) is by plotting 
$\frac{F_{\rm sessile}}{F_{\rm pendant}}$ (see Fig.~\ref{fig:sop}) and 
$\frac{F_{\rm inverted}}{F_{\rm sessile}}$ (see Fig.~\ref{fig:ios}), which
according to Eq.~(\ref{seqoffor}) should be greater than one. It is clear from 
Fig.~\ref{fig:sop} that $F_{\rm sessile}$ is always greater than 
${F_{\rm pendant}}$. It is also clear that as the 
volume decreases, $F_{\rm sessile}$ and ${F_{\rm pendant}}$ become closer to each 
other. Indeed, for 100-$\mu$L drops, $F_{\rm sessile}$ is
18.9\% larger than ${F_{\rm pendant}}$, but for 15-$\mu$L drops, 
$F_{\rm sessile}$ is only 2.4\% larger than ${F_{\rm pendant}}$. This is not
surprising, because the influence of gravity compared to that of surface
tension decreases as the volume decreases, and therefore the drops become 
more alike as they become smaller.

We can see in Fig.~\ref{fig:ios} that $F_{\rm inverted}$ is comparable
to, and usually larger than, $F_{\rm sessile}$. We can also see that, unlike
Fig.~\ref{fig:sop}, Fig.~\ref{fig:ios} does not show a steady decrease
of the ratio $\frac{F_{\rm inverted}}{F_{\rm sessile}}$ as the volume
decreases. What is more,
for 25~$\mu$L, $F_{\rm inverted}$ is essentially the same as $F_{\rm sessile}$. The
reason is that it is difficult to prepare inverted sessile drops by flipping
a surface. In fact, unless it is done properly~\cite{AJP15}, the drop 
tends to slide on the surface during the flip, which may deform the drop and
make it loose some of its adhesion to the surface.

Another way to rephrase Eq.~(\ref{seqoffor}) is by comparing the 
centrifugal accelerations at the onset of the motion, see 
Table~\ref{table:centriacc}.
As can be seen in Table~\ref{table:centriacc}, sessile drops start sliding at
a larger acceleration than pendant drops, but at a smaller 
acceleration than inverted 
drops, and hence Eq.~(\ref{seqoffor}) holds.

\vskip1cm

\begin{table}[!h]
\begin{center}
\begin{tabular}{l| c c c c c c c c c c c} 
\hline\hline
\multicolumn{1}{c}{\quad} & \multicolumn{10}{c}{Centrifugal Acceleration 
            (m/s$^2$)} \\
\hline

{\bf Volume} ($\mu$L)  & 100 & 90 & 80 & 70 & 	60 &	 50 &	40 &	
          30 & 25 &  20 & 15  \\
\hline
{\bf Sessile} & 2.75 & 3.01 &  3.13 & 3.50 &  3.89 & 4.62 & 5.24 & 6.50 & 7.31
    & 8.56 &   10.22  \\

{\bf Pendant} & 2.32 & 2.56 & 2.71 & 3.07 &  3.49 & 4.06 & 4.72 & 6.16 & 6.89
& 8.17 & 9.97 \\

{\bf Inverted} & & & & & & & &  6.68 & 7.39 & 8.84  &  10.56 \\

\hline\hline 
\end{tabular}
\caption{Centrifugal accelerations at the onset of motion. Statistical errors
are about 2\% or less.}
\label{table:centriacc}
\end{center}
\end{table}

\subsection{Systematics, troubleshooting, and reproducibility}

To exclude systematic effects as the
source of our results, we performed two additional experiments. One experiment
was similar to that of Ref.~\cite{AJP15}, and it yielded the same 
results as the present paper. The other one, which can be very easily 
reproduced, was a simple tilted-plate experiment. We placed a sessile and a 
pendant drop on a PMMA surface and then slowly tilted the surface until the 
drops started to slide. We visually observed that, on average, the pendant 
drops started to slide before the sessile drops. However, when the same 
experiment was done with sessile and inverted sessile drops, we observed 
that, on average, the sessile drops started to slide before the inverted 
sessile drops. We also did the tilted-plate experiment with water drops 
on polycarbonate (Lexan) and obtained the same results.

The most important factor affecting the retention force
in our experiments is how the drop is formed. While preparing the drops,
the pipette needs to be perpendicular to the 
surface and above the same point of the surface. Otherwise, the shape and the
width of the drop may change, which will affect the retention force 
of the drop. In addition, the procedure of Ref.~\cite{AJP15} to build 
inverted drops must be followed. Overall, it is not 
difficult to build sessile and pendant drops that are consistently similar,
but inverted sessile drops require some skill and practice.

\subsection{Derivation of the theoretical retention force}
\label{sec:derivation}

To really understand our theoretical explanation of the 
above experimental observations, it is useful to first 
understand the derivation of the theoretical retention 
force~\cite{MACDOUGALL,FRENKEL,KAWASAKI,
FURMIDGE,BROWN,DUSSAN,EXTRAND,CARRE,ELSHERBINI,CONINCK}. In this 
section, we provide such derivation. Our derivation is an
improved version of the derivation provided by Dussan and Chow~\cite{DUSSAN}.

\begin{figure}[h!]
\begin{center}
              \epsfxsize=10cm
              \epsffile{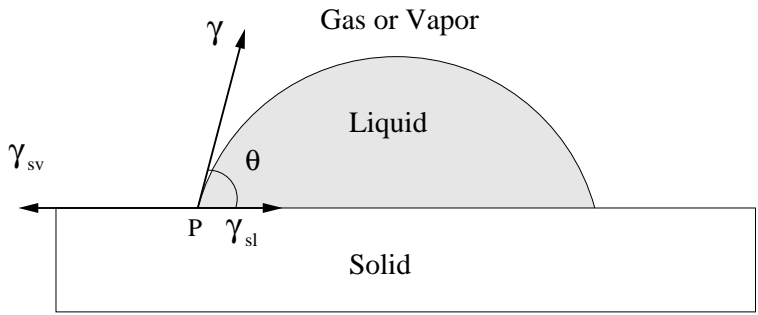}
\end{center}                
\caption{Surface tensions and contact angle at P. Point P is
a generic point along the triple line, not necessarily the advancing
or receding edge.}
\label{fig:stcca}
\end{figure}

Let us consider a drop on a flat surface subject to an external 
force parallel to the surface. Three surface tensions act on a given
infinitesimal section $ds$ of the triple line at point P 
(see Fig.~\ref{fig:stcca}):
The solid-liquid surface tension $\gamma_{\rm sl}$,
the solid-vapor surface tension $\gamma _{\rm sv}$, and the liquid-vapor
surface tension $\gamma \equiv \gamma _{\rm lv}$.

\begin{figure}[h!]
\begin{center}
              \epsfxsize=10cm
              \epsffile{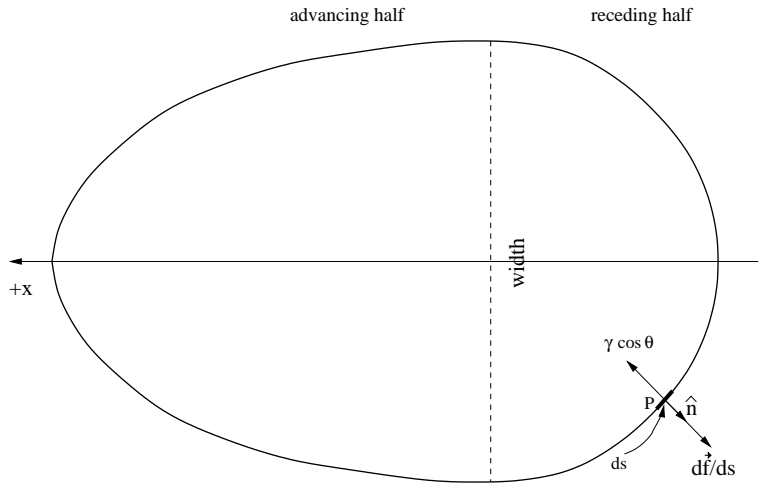}
\end{center}                
\caption{Triple line and contact area of a drop at the onset 
of lateral motion. The
centrifugal force points in the $+x$-direction. The width $w$
is the length of the dashed line.}
\label{fig:contacta}
\end{figure}

The solid-liquid and solid-vapor surface tensions act parallel to the surface, 
and the liquid-vapor
surface tension acts at an angle $\theta$ with respect to the surface, where
$\theta$ is the contact angle at that point. Thus, in the direction
parallel to the surface but perpendicular to the triple line, the surface 
tensions acting on an element $ds$ of triple line located at P are
$\gamma \cos \theta$, 
$\gamma_{\rm sl}$ and $\gamma _{\rm sv}$. When the contact angle is the Young,
equilibrium angle $\theta_{\rm Y}$, the net force on the element $ds$
is zero, which
leads to the Young equation:
\begin{equation}
       \gamma \cos \theta_{\rm Y} +\gamma_{\rm sl}-\gamma_{\rm sv}=0 \, .
\end{equation}
When the contact angle $\theta$ at $ds$ is not $\theta _{\rm Y}$, the
forces due to the surface tensions do not cancel each other,
\begin{equation}
       \gamma \cos \theta + \gamma_{\rm sl}-\gamma_{\rm sv} \neq 0 \, ,
        \label{sost}
\end{equation}
and therefore there is a nonzero capillary force per unit of length
acting on the element $ds$ in the direction
parallel to the surface. If we denote by $\nhat$ the unit vector
perpendicular to the triple line and pointing outwardly (see
Fig.~\ref{fig:contacta}), then such
capillary force per unit of length is given by
\begin{equation}
       (-\gamma \cos \theta -\gamma_{\rm sl}+\gamma_{\rm sv})\, \nhat \, .
        \label{cfpuol}
\end{equation}
However, if the infinitesimal element $ds$ at point P does not move,
there must be a force that cancels the force in Eq.~(\ref{cfpuol}), much 
like when we push a solid resting on a surface but the solid doesn't move, we 
say that our push is canceled by the static frictional force. The force that 
cancels that in Eq.~(\ref{cfpuol}) is the retention 
force per unit of length and is presumably due to solid-liquid-vapor
interactions (compare with Eq.~(6.1) in Ref.~\cite{DEGENNES},
Eq.~(4.1) in Ref.~\cite{BORMASHENKO}, and Eqs.~(11) and~(12) in
Ref.~\cite{CARRE}),
\begin{equation}
    \frac{d\fp}{ds} = 
           (\gamma \cos \theta +\gamma_{\rm sl}-\gamma_{\rm sv})\, \nhat \, .
        \label{magretforvec}
\end{equation}
Figure~\ref{fig:contacta} shows the direction of the retention force at the
infinitesimal element of triple line located at point P.

The net retention force on the whole drop is obtained by adding the
infinitesimal retention forces acting on each infinitesimal element of the
triple line $C$:
\begin{equation}
    \fp = \oint_C 
           (\gamma \cos \theta +\gamma_{\rm sl}-\gamma_{\rm sv})\, \nhat\, ds \, .
        \label{netforcepf}
\end{equation}
Because $\gamma$, $\gamma_{\rm sl}$ and $\gamma_{\rm sv}$ are 
constant~\cite{NOTE1}, and 
because $\oint_C \nhat\, ds =0$, Eq.~(\ref{netforcepf}) 
simplifies to
\begin{equation}
    \fp = \gamma \oint_C \cos \theta\, \nhat \, ds \, .
        \label{netforce}
\end{equation}
This is the exact equation that provides the retention force on a
drop resting on a uniform solid. When the contact area is 
symmetric with respect to the $x$-axis, the $y$-component of this 
force is zero, and therefore one only needs to worry about the $x$-component,
\begin{equation}
    \fpm = {\fpm}_x  = \gamma \oint_C \cos \theta\, \nhat \cdot \xhat \, ds \, .
        \label{netforcex}
\end{equation}
To obtain the exact expression of $\fpm$, one needs to know 
\begin{itemize}
    \item[({\it i})] the exact shape of the triple line and, in particular, 
the variation along $C$ of the angle between 
$\nhat$ and the centrifugal force, i.e., the variation of 
$\nhat \cdot \xhat$ along the triple line,
    \item[({\it ii})] how the
contact angle $\theta$ varies along the triple line.
\end{itemize}

Because it is very difficult to obtain ({\it i})-({\it ii}) either 
experimentally or theoretically, it is common to make some approximations 
to obtain an effective, approximate expression for the retention force 
that captures the essence of the exact one. One common approximation 
is to assume that in the ``advancing half''
(``receding half'') of the triple line, the contact angle remains constant 
and equal to the contact angle at the advancing (receding) edge of the drop, 
$\theta _{\rm a}$ ($\theta _{\rm r}$). Within this approximation, and denoting by
$C_{\rm a}$ ($C_{\rm r}$) the contour associated with the
advancing (receding) half of the triple line, Eq.~(\ref{netforcex})
becomes
\begin{eqnarray}
   \fpm &=& \gamma \left( \int_{C_{\rm a}} \cos \theta \, \nhat \cdot \xhat \, ds +
           \int_{C_{\rm r}} \cos \theta \, \nhat \cdot \xhat \, ds \right) 
              \label{1step}  \\
      & \approx &
         \gamma \left( \cos \theta _{\rm a} \int_{C_{\rm a}} 
          \nhat \cdot \xhat\, ds + \cos \theta _{\rm r}
           \int_{C_{\rm r}} \nhat \cdot \xhat \, ds \right)  \label{2step}
            \\
     &=& \gamma w \left( \cos \theta _{\rm r}- \cos \theta _{\rm a} \right) 
        \label{netforcexa}
\end{eqnarray}
where in the last step we have used~\cite{DUSSAN,CONINCK} that
$\int_{C_{\rm r}} \nhat \cdot \xhat\, ds=-\int_{C_{\rm a}} \nhat \cdot \xhat\, ds=w$,
$w$ being the width of the drop in the direction perpendicular to
the motion of the drop, see Fig.~\ref{fig:contacta}. 

It is clear that, in going from Eq.~(\ref{1step}) to Eq.~(\ref{2step}), 
we are making an approximation. It is also clear that
what we call the ``advancing'' and ``receding'' halves is somewhat
arbitrary. To correct for these approximations and still have
a useful formula, it is common to introduce a shape factor $k$ in the
retention force,
\begin{equation}
      \fpm = k \gamma w \left( \cos \theta _{\rm r}- \cos \theta _{\rm a} \right)
        \, .
   \label{fpw}
\end{equation}
The shape factor $k$ carries the information lost in the approximation, 
that is, $k$ accounts for ({\it i})-({\it ii}). From
a practical point of view, the shape factor is simply a fitting parameter
that makes the theoretical retention force $\fpm$ be equal to the
experimental one $F$. Hence, one can calculate $k$ in terms of experimentally
measurable quantities as
\begin{equation}
   k= \frac{F}{\gamma w \left( \cos \theta _{\rm r}- \cos \theta _{\rm a} \right)}
    \, .
    \label{kinterofex}
\end{equation}

One can improve the above approximation by assuming a specific variation of the
contact angle $\theta$ along the triple line. This is what, for example,
Extrand and Gent~\cite{EXTRAND} and ElSherbini and Jacobi~\cite{ELSHERBINI}
did, who obtained $k=2/\pi$ and $k=24/\pi^3$, respectively. It
should be noted however that Refs.~\cite{EXTRAND,ELSHERBINI} assumed
that the angle between $\xhat$ and $\nhat$ is the same as the polar angle 
that locates each point on the triple line. Using an argument similar
to ours, Carre and Shanahan derived Eq.~(\ref{fpw}) with a shape
factor of $\pi/4$.

To finish this section, we would like to note that the advancing (receding)
angle is usually defined as the angle that the
drop makes with the surface when the drop is inflated 
(deflated)~\cite{DEGENNES,BORMASHENKO}. However, there are situations in
which the contact angle at the advancing (receding) edge of the drop
is not the same as when the drop is inflated (deflated)~\cite{MARMUR}. In our
case, due to gravity, the contact angles at the advancing and receding edges 
of a sessile drop are not the same as those of pendant and inverted 
drops. However, as the above derivation shows, this is of no 
concern: Equation~(\ref{fpw}) simply uses the angle at the advancing 
($\theta _{\rm a}$) and receding ($\theta _{\rm b}$) edges of the drop for the
situation at hand, independently of whether such angles are different in
other situations.

\subsection{Theoretical explanation}

According to Eq.~(\ref{fpw}), the three main factors that 
affect the retention force are the shape of the drop ($k$), its width ($w$),
and contact angle hysteresis ($\cos \theta _{\rm r}-\cos \theta _{\rm a}$). In
this section, we are going to elucidate the role that each of these 
factors plays in differentiating the retention forces on sessile,
pendant, and inverted drops.

To elucidate whether the width of the drop plays
an essential role in the difference between
$F_{\rm sessile}$, $F_{\rm pendant}$ and $F_{\rm inverted}$, we calculated the 
centrifugal force per unit of width,
\begin{equation}
   \frac{F}{w}= \left\{ \begin{array}{cll}
                           (30.1 \pm 1.2)~{\rm mN}/{\rm m} & \hskip0.5cm  &  \text{sessile,}
                           \ \operatorname{15-100}~\mu \text{L} , \\
                           (29.9 \pm 0.9)~{\rm mN}/{\rm m} &  &\text{pendant,} \ 
                                  \operatorname{15-100}~\mu \text{L} , \\
 (31.2 \pm 0.9)~{\rm mN}/{\rm m} & \hskip0.5cm  &  \text{inverted,} \
                        \operatorname{15-30}~\mu \text{L} . \\
                               \end{array}     \right.
              \label{fowt}
\end{equation}
Because $F/w$ is essentially the same for sessile, pendant and inverted 
drops~\cite{NOTE4}, we conclude that the retention force is proportional 
to the width of the drop. Thus, since for each volume
$w_{\rm pendant}<w_{\rm sessile}<w_{\rm inverted}$ (see Table~\ref{table:finalw}), 
it follows that $F_{\rm pendant}<F_{\rm sessile}<F_{\rm inverted}$.

\vskip1cm

\begin{table}[!h]
\begin{center}
\begin{tabular}{l| c c c c c c c c c c c} 
\hline\hline
\multicolumn{1}{c}{\quad} & \multicolumn{10}{c}{Width (mm) } \\
\hline

{\bf Volume} ($\mu$L)  & 100 & 90 & 80 & 70 & 	60 &	 50 &	40 &	
          30 & 25 &  20 & 15  \\
\hline
{\bf Sessile} & 9.48 & 9.15 & 8.80 & 8.51 & 8.04 & 7.83 & 6.75 & 6.18
    & 5.83 & 5.44 &  4.90 \\

{\bf Pendant} & 8.01 & 7.64 & 7.46 & 7.23 & 7.21 & 6.90 & 6.39 & 5.90
 & 5.58 & 5.32 & 4.85  \\

{\bf Inverted} & & & & & & & &  6.21 & 5.89 & 5.86 & 5.09 \\

\hline\hline 
\end{tabular}
\caption{Widths of the drops at the onset of motion. Statistical errors
are at the most 0.6\%.}
\label{table:finalw}
\end{center}
\end{table}

To determine the influence of contact angle hysteresis on each type of
drop, we measured the 
contact angles at the 
advancing and receding edges, and we obtained that
$\cos \theta _{\rm r}-\cos \theta _{\rm a}$ is very similar for sessile,
pendant, and inverted drops,
\begin{equation}
   \cos \theta _{\rm r}-\cos \theta _{\rm a} = \left\{ \begin{array}{cll}
                           0.419 \pm 0.017 & \hskip0.5cm  &  \text{sessile,} \\
                           0.426 \pm 0.013 &  &\text{pendant,} \\
                           0.428 \pm 0.007 &  &\text{inverted.} \\
                               \end{array}     \right.
           \label{cahis}
\end{equation}
This result, coupled to the fact that $\theta _{\rm r}$ and $\theta _{\rm a}$ 
remain fairly constant as the volume of the drops decreases, leads us to 
conclude that it is unlikely that contact angle hysteresis is essential in 
differentiating the retention forces on sessile, pendant and inverted 
drops~\cite{NOTE3}.

As mentioned above, the shape factor $k$ is, for
practical purposes, a free parameter that allows one to fit the
theoretical retention force $\fpm$ to the experimental one $F$. Thus, it
is difficult to determine the influence of the shape of the drop on the
retention force. One can nevertheless use Eq.~(\ref{kinterofex}) to obtain
a value of the shape factor in terms of experimentally measurable 
quantities. In our experiment, we obtained that 
\begin{equation}
   k = \left\{ \begin{array}{cll}
                           0.98 \pm 0.08 & \hskip0.5cm  &  \text{sessile,} \\
                           0.97 \pm 0.04 &  &\text{pendant,} \\
                           1.00 \pm 0.05 &  &\text{inverted.} \\
                               \end{array}     \right.
           \label{sfspi}
\end{equation}
Because these shape factors are fairly similar, it is unlikely that the shape 
of the triple line and the variation of the contact angle along such line are
critical in determining which type of drop gets more pinned to the surface.

Thus, among the three factors that make up
the retention force in Eq.~(\ref{fpw}), the width seems to be the one that
most critically determines which kind of drop gets more pinned to the 
surface. Because the width of a sessile
drop is larger than the width of a pendant drop 
(see Table~\ref{table:finalw}), and because the retention force is 
proportional to the width, we have that 
$F_{\rm sessile}$ is larger than $F_{\rm pendant}$. However, when we flip a 
sessile drop and make it an inverted sessile drop, at the onset
of the motion the width is slightly larger than
what it would had been for the original sessile drop (as long as the flip 
is done as explained in Ref.~\cite{AJP15}), and that is what makes
$F_{\rm inverted}$ larger than $F_{\rm sessile}$.

One can prepare drops of a given type and volume, e.g., 80-$\mu$L sessile
drops, in different ways so that
the widths of the drops are different. Because it is proportional
to the width, the retention force should be larger for those drops with a
larger width,
even though they all are sessile drops of the same volume. Indeed, an 
experiment was done in Ref.~\cite{AJP15} with 80-$\mu$L sessile
drops that were formed with different widths relative to the radial direction of
the motion. It was found that the retention force increased with the width,
even though the volume and the type of drop were the same. That the
experiment in Ref.~\cite{AJP15} with 80-$\mu$L drops can be easily
explained by saying that the retention force is proportional to the width 
further suggests that the width is the most critical parameter in 
differentiating $F_{\rm sessile}$, $F_{\rm pendant}$ and $F_{\rm inverted}$, and that
the retention force depends not only on whether
the drop is placed on top or bottom of a solid surface, but also on the history
of droplet deposition.

The lack of volume dependence in Fig.~\ref{fig:ios} can also be understood
from the proportionality between the retention force and the width. Due to 
the way we build inverted drops, the ratio
$\frac{w_{\rm inverted}}{w_{\rm sessile}}$ is close to unity for any volume, and 
therefore $\frac{F_{\rm inverted}}{F_{\rm sessile}}$ should always be close to unity, 
independently of the volume of the drop.

It is important to mention that the influence of the history of drop 
deposition on the retention force has been studied recently by
R\'\i os-L\'opez {\it et al.}~\cite{GRIEGOS2,GRIEGOS3}, who found that one
can still use Eq.~(\ref{fpw}) even when the drop history contains an initial 
stage of tilting. In their case, the relevant geometrical feature that
enters Eq.~(\ref{fpw}) is the initial length of the drop, whereas in
our case it is the width at the onset of the motion.

\section{Conclusions}

Using a rotating-platform experiment, we have observed that
the lateral retention force on sessile drops is larger than on pendant
drops, but smaller than on inverted sessile drops. We have identified
the width of the drop as the critical parameter that determines this
result, even though the contact angles and the shape of the triple line
also affect the retention force. Because the retention force
is proportional to the width, and because we built the drops such that
$w_{\rm pendant}<w_{\rm sessile}<w_{\rm inverted}$, we have that
$F_{\rm pendant}<F_{\rm sessile}<F_{\rm inverted}$. In general, the retention force
on a drop does not depend only on whether it is placed on top or on bottom
of a surface, but also on how the drop was formed. Our results can be 
easily reproduced, by using either a simple tilted-plate experiment or other
drop accelerators~\cite{TADMOR09,GRIEGOS1}.

We would like to note that the experiment in Ref.~\cite{TADMOR09}
was done with a different liquid-solid combination, and for smaller 
volumes. We believe that our results would also be applicable to many
other liquid-solid combinations for large volumes. However, for very small
volumes, it might be
possible that $F_{\rm sessile}$ is smaller than $F_{\rm pendant}$, as long as
$\cos \theta _{\rm r}-\cos \theta _{\rm a}$ remains constant as the volume of
the drops decreases~\cite{NOTE3}. If such was the case, then contact angle
hysteresis should suffice to explain why $F_{\rm sessile}$ becomes smaller
than $F_{\rm pendant}$. Our experiment, however, does not address the effect 
of the resting time on the retention force~\cite{TADMOR09}.

\section*{Acknowledgments}

The authors thank Rafael Tadmor for many enlightening 
discussions. Additional thanks are due to Sage Automation, Thomas Michel, 
Thomas Podgorski, Taylor Whitehead, Jason Dark, Jon Klipfel, Aaron
Burlew, David Jackson, Miles Stone, Jared Richards, 
and Al Sauerman. Financial support from a Lamar Presidential Fellowship 
is gratefully acknowledged.

\newpage

\begin{figure}[h!]
\begin{center}
              \epsfxsize=15cm
              \epsffile{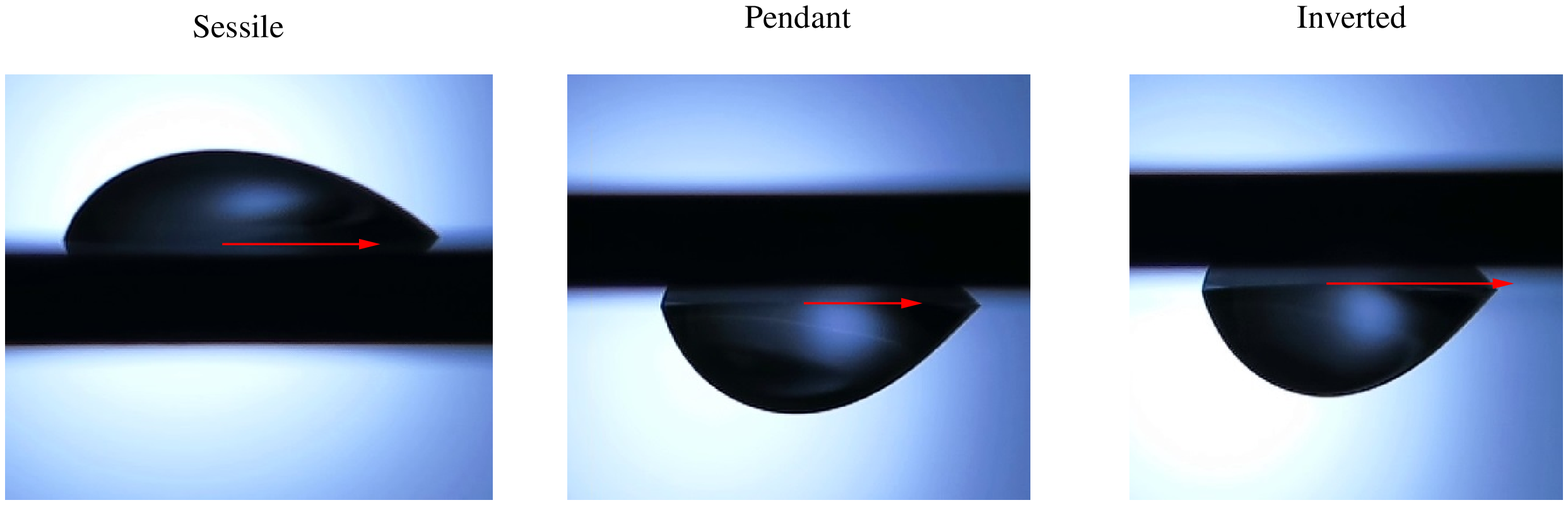}
\end{center}                
\caption{{\bf Table of Contents Graphic:} 
30-$\mu$L sessile, pendant and inverted 
drops at the moment they start sliding on the PMMA sheet. The red arrow 
represents the lateral retention force (not to scale).}
\label{fig:tocg}
\end{figure}


\begin{thebibliography}{99}
\bibitem{SYRINGE} Even though syringes are slightly more precise than
micropipettes, we did not use them. The reason is that
when we built pendant drops with a syringe, there was 
more liquid left in the syringe as compared to when we built sessile and 
inverted drops. With a micropipette, you can visually check that all 
the liquid has been deposited on the surface.
\bibitem{VIBRATION} As the angular velocity increases, small vibrations
of the apparatus produce a small trembling motion on surface of the drops. The 
triple line however does not seem to be significantly affected by such 
vibrations.
\bibitem{NOTE1} If $\gamma_{\rm sl}$ and $\gamma_{\rm sv}$ were not 
constant along the triple line (for example, if the drop straddled the
boundary between two different solids), then 
$\oint_C (\gamma_{\rm sl}-\gamma_{\rm sv}) \nhat\, ds$ would not be zero, and
the retention force would be provided by Eq.~(\ref{netforcepf}),
not by Eq.~(\ref{netforce}).
\bibitem{NOTE4} Although it remains fairly constant, $\frac{F}{w}$ has a 
slight tendency to increase as the volume decreases.
\bibitem{NOTE3} Due to gravity and the deformability of
liquid water, the contact angles of sessile drops 
are about 6$^\circ$ smaller than those of pendant and inverted drops. Thus,
$\cos \theta _{\rm r}-\cos \theta _{\rm a}$ is always a bit smaller for
sessile drops. Such difference in 
$\cos \theta _{\rm r}-\cos \theta _{\rm a}$ is however not enough 
to upset Eq.~(\ref{seqoffor}), at least in the volume range we have
studied.
\end{thebibliography}
\end{document}